\documentclass[useAMS,usenatbib]{mn2e}
\usepackage{graphicx} 
\usepackage{amsmath} 
\usepackage{color}
\usepackage{caption}
\usepackage{placeins}
\usepackage{hyperref}
\usepackage{cleveref}
\usepackage{subcaption}
\usepackage{float}
\usepackage{lipsum}
\usepackage{multicol}

\usepackage{amssymb}
\def\be{\begin{equation}} 
\def\ee{\end{equation}}

\def\msun{{\Msun}}

\def\HI{\hbox{H~$\scriptstyle\rm I$}}

\def\gsim{\lower.5ex\hbox{\gtsima}} 
\def\lsim{\lower.5ex\hbox{\ltsima}} \def\gtsima{$\; \buildrel > \over 
\sim \;$} \def\ltsima{$\; \buildrel < \over \sim \;$} \def\prosima{$\; 
\buildrel \propto \over \sim \;$} \def\gsim{\lower.5ex\hbox{\gtsima}} 
\def\lsim{\lower.5ex\hbox{\ltsima}} 
\def\simgt{\lower.5ex\hbox{\gtsima}} 
\def\simlt{\lower.5ex\hbox{\ltsima}} 
\def\simpr{\lower.5ex\hbox{\prosima}} \def\la{\lsim} \def\ga{\gsim} 
  
 \def\gtsima{$\; \buildrel > \over \sim \;$} 
\def\ltsima{$\; \buildrel < \over \sim \;$} 
\def\gsim{\lower.5ex\hbox{\gtsima}} 
\def\lsim{\lower.5ex\hbox{\ltsima}} 
\def\simgt{\lower.5ex\hbox{\gtsima}} 
\def\simlt{\lower.5ex\hbox{\ltsima}} 
\def\simpr{\lower.5ex\hbox{\prosima}} 
\def\la{\lsim} 
\def\ga{\gsim}

\def\msun{\,{\rm \Msun}}
\def\mstar{\,{\rm \dot M_*}}

\def\E3{{\cal E}_{\rm g}^{III}}

\def\Msun{\rm M_\odot}

\def\Msun{\rm M_\odot}

\def\M*{M_*}

\def\Z*{Z_*}
\def\L*{L_*}

\def\der{{\rm d}} 

\def\kev{\rm keV}
\def\mx{\,m_x}

\def\der{{\rm d}}

\newcommand{\fig}[1]{Figure~\ref{#1}}

\title[Ruling out 3 keV WDM using EDGES]{Ruling out 3 keV warm dark matter using 21~cm-EDGES data} 
\author[Chatterjee, Dayal, Choudhury \& Hutter]{Atrideb Chatterjee$^{1}$\thanks{atrideb@ncra.tifr.res.in}, 
Pratika Dayal$^{2}$,
Tirthankar Roy Choudhury$^{1}$ \& 
Anne Hutter$^2$ \\ 
$^{1}$ National Centre for Radio Astrophysics, Tata Institute of Fundamental Research, Pune 411007, India\\
$^{2}$ Kapteyn Astronomical Institute, University of Groningen, P.O. Box 800, 9700 AV Groningen, The Netherlands
}
\begin{document} 
 
\date{} 

\maketitle

\begin{abstract}
Weakly interacting cold dark matter (CDM) particles, which are otherwise extremely successful in explaining various cosmological observations, exhibit a number of problems on small scales. One possible way of solving these problems is to invoke (so-called) warm dark matter (WDM) particles with masses $m_x \sim$ keV. Since the formation of structure is delayed in such WDM models, it is natural to expect that they can be constrained using observations related to the first stars, e.g., the 21~cm signal from cosmic dawn. In this work, we use a detailed galaxy formation model, {\it Delphi}, to calculate the 21~cm signal at high-redshifts and compare this to the recent EDGES observations. We find that while CDM and 5 keV WDM models can obtain a 21~cm signal within the observed redshift range, reproducing the amplitude of the observations requires the introduction of an excess radio background. On the other hand, WDM models with $m_x \lsim 3$ keV can be {\it ruled out} since they are unable to match either the redshift range or the amplitude of the EDGES signal, irrespective of the parameters used. Comparable to values obtained from the low-redshift Lyman Alpha forest, our results extend constraints on the WDM particle to an era inaccessible by any other means; additional forthcoming 21~cm data from the era of cosmic dawn will be crucial in refining such constraints.

\end{abstract}

\begin{keywords}
cosmology: dark ages, reionization, first stars -- theory - dark matter; galaxies: formation -- intergalactic medium
\end{keywords}

\section{Introduction}
The standard cold dark matter (CDM) paradigm, where structure formation is driven by cold and weakly interacting dark matter particles has been verified at scales ranging from the Lyman-Alpha (Ly$\alpha$) forest to the large scale structure ($10-100$ Mpc) of the Universe \citep[for a review see][]{weinberg2015}. However CDM exhibits a number of small scales problems \citep[for a recent review see, e.g.,][]{dayal2018} that include the observed lack of both theoretically predicted low and high-mass satellites of the Milky Way \citep{moore1999, klypin1999,boylan2011,boylan2012}, dark matter halos forming high-density (cuspy) centres compared to the observationally preferred constant density cores \citep{moore1999b, subramanian2000} and encountering difficulties in producing typical disks due to mergers down to redshifts as low as $z \simeq 1$ \citep{wyse2001}. Given the limited success of baryonic feedback in solving these issues \citep[e.g.][]{boylan2012, teyssier2013}, a growing body of work has focused on questioning the (cold) nature of dark matter itself. Alternatives to CDM include Warm Dark Matter (WDM) with particle masses $m_x \sim \mathcal{O}$(keV) \citep[e.g.][]{blumenthal1984,bode2001}, fuzzy CDM with $m_x \sim \mathcal{O}(10 ^{-22}$eV) \citep{hui2017,du2017}, self-interacting (1 MeV - 10 GeV) dark matter \citep{spergel2000, vogelsberger2014_wdm}, decaying dark matter \citep{wang2014} and interactive CDM \citep[e.g.][]{boehm2001, dvorkin2014}, to name a few. 

Recently, the EDGES (Experiment to Detect the Global Epoch of Reionization Signature) collaboration has measured a sky-averaged absorption signal at a central frequency $\nu = 78\pm 1$ Mhz corresponding to $z \approx 17.2$ \citep{bowman2018}. While this redshift is consistent with CDM expectations of Ly$\alpha$ photons coupling the 21~cm brightness temperature to the gas kinetic temperature, the differential brightness temperature $\sim -500\pm 200$mK is about twice as strong as that expected from any ``standard" model \citep{barkana2018}. Explaining the strength of this signal either requires gas colder than expected or a radiation background higher than expected \citep{bowman2018}. Indeed, if verified, this signal either points to new dark matter physics \citep{barkana2018, fraser2018, pospelov2018, slatyer2018}, an excess radio background \citep{feng2018, fialkov2019} from sources ranging from a population of early black holes \citep{ewall_wice2018} to black hole-high-mass X-ray binary micro-quasars \citep{mirabel2019} to Population III (PopIII; metal-free) stars \citep{jana2019}, an early Ly$\alpha$ background from PopIII stars \citep{schauer2019} or extremely efficient star formation in low-mass ($10^{8-9}\msun$) halos \citep{mirocha2019}. 

The redshift and strength of this signal allow an ideal opportunity to obtain constraints on the allowed WDM particle mass {\it at redshifts currently unaccessible by any other means}. This is because WDM models with low $m_x$ values will find it increasingly hard to explain the signal given the lack of low-mass structures at such high-$z$. In this work, we focus on four dark matter models: CDM and WDM with $m_x=1.5$, 3 and 5 keV. While similar in spirit to \citet{safarzadeh2018}, our model is more advanced in that it follows the joint assembly of both dark matter and baryons, properly accounting for feedback and the cosmology-dependent values of ionizing photon and Lyman-Werner (LW) photon production, in each model as detailed in what follows. 

The cosmological parameters used in this work correspond to ($\Omega_m,\Omega_{\Lambda}, \Omega_b, h, n_s, \sigma_8) = (0.3089,0.6911,0.049, 0.67, 0.96, 0.81)$, consistent with the latest results from the {\it Planck} collaboration \citep{planck2015}.

\section{The theoretical model}
We start by summarising the galaxy formation model used before discussing how it is used to infer the 21~cm brightness temperature.

\subsection{The galaxy formation model: Delphi}
We use the {\it Delphi} ({\bf D}ark Matter and the {\bf e}mergence of ga{\bf l}axies in the e{\bf p}oc{\bf h} of re{\bf i}onization) code, introduced in \citet{dayal2014, dayal2015}, to track the build-up of dark matter halos and their baryonic component (both gas and stellar mass) over the first billion years. We start with 400 (800) $z=4$ galaxies for CDM and WDM models with $\mx = 3$ and $5\, \kev$ (WDM with $\mx = 1.5\, \kev$), linearly distributed across the halo mass range $\log(M_h/\Msun) = 9-13$, with a mass resolution of $10^8 \Msun$. We construct merger trees for these galaxies, over 320 equal redshift steps between $z=20$ and $4$, using the modified binary merger tree algorithm with smooth accretion detailed in \citet{parkinson2008} and \citet{benson2013}. As detailed in \citet{dayal2015}, our WDM merger trees are computed according to the prescription in \citet{benson2013} wherein the authors show that a number of modifications have to be introduced to obtain WDM merger trees and mass functions that are in agreement with N-body simulations. These include using a modified initial power spectrum that imposes a cut-off in power below a certain length scale depending on the WDM particle mass, using a critical over-density for collapse that depends on the WDM particle mass, using a sharp window function in {\it k}-space and calibrating the smooth-accretion of DM from N-body simulations. As shown in Fig. 1 \citet{dayal2015}, this induces a turn-over in the 1.5 keV WDM halo mass function at about $10^{9.5}$ ($10^{10}$) solar masses at $z \sim 7$ (12). Each $z=4$ halo is assigned a co-moving number density by matching to the $\der n / \der M_h$ value of the $z=4$ Sheth-Tormen halo mass function (HMF) and every progenitor halo is assigned the number density of its $z=4$ parent halo; the resulting HMFs are found to be in good agreement with the Sheth-Tormen HMF at all redshifts\footnote{While the Sheth-Tormen HMF provides a fast and analytical way to generate the merger trees, we remind the readers that they are found to be \emph{not} in agreement with simulations at high redshifts \citep{watson2013,trac2015}. This may affect our conclusions to some extent, however, quantifying this effect is difficult without detailed $N$-body simulations and hence beyond the scope of this paper.}. 

The very first progenitors (starting leaves) of any halo are assigned an initial gas mass that scales with the halo mass such that $M_g = (\Omega_b/\Omega_m) M_h$. The fraction of this gas mass that can form stars depends on the effective star formation efficiency, $f_*^{eff}$, of the host halo. The value of $f_*^{eff}$ for any halo is calculated as the minimum between the star formation efficiency that produces enough type II supernova (SNII) energy to eject the rest of the gas, $f_*^{ej}$, and a maximum threshold, $f_*$, so that $f_*^{eff} = min[f_*^{ej}, f_*]$. The stellar mass produced at any $z$-step is calculated as $M_*(z) = f_*^{eff} M_g(z)$. The final gas mass, at the end of that $z$-step is given by $M_{gf}(z) = [M_g(z)-M_*(z)] [1-(f_*^{eff}/f_*^{ej})]$. At each $z$-step we also account for smooth accretion of dark matter from the inter-galactic medium (IGM) and reasonably assume this to be accompanied by smooth-accretion of a cosmological fraction of gas mass. Throughout this work, we use a Salpeter initial mass function \citep[IMF;][]{salpeter1955} between $0.1-100\Msun$. 

Over the past years, our group has carried out extensive calculations to successfully confront this framework with all available observational data-sets for high-$z$ galaxies \citep{dayal2015}, reionization \citep{dayal2017a} and the IGM metal enrichment at high-$z$ \citep{bremer2018} in both CDM and WDM cosmologies for particle masses ranging between $m_x = 1.5-5$ keV. We find that matching to observations of high-$z$ galaxies requires roughly 10\% of the SNII energy to couple to gas and a maximum (instantaneous) star formation efficiency of $f_* = 3\%$; these are the only two mass- and $z$-independent free parameters used in the galaxy formation model.  We then use the full stellar mass assembly history from this data-benchmarked model to obtain the spectrum for each galaxy using the population synthesis code Starburst99 \citep{leitherer1999}; from this we infer the total output of ionizing photons and, hence, the Ly$\alpha$ luminosity produced by any given halo. Further, the Ultra-violet (UV) luminosity obtained from the model is converted into a star formation rate (SFR; $\mstar$) using the relation ${\rm L_{UV}} = 7 \times 10^{27} ({\rm \mstar}/\msun\, {\rm yr^{-1}}) [{\rm erg\, s^{-1} \, Hz^{-1}}]$ appropriate for our chosen IMF and in excellent agreement with the values generally used \citep[from e.g.][]{madau1998}. 

Now that the underlying galaxy model has been detailed, we discuss the calculation of the 21~cm differential brightness temperature in what follows. 
\subsection{The 21cm differential brightness temperature}

Our calculation of the 21~cm signal from cosmic dawn closely follows that of \citet{furlanetto2006c} and \citet{pritchard2012}. We summarise the most salient points of the calculation in this section and interested readers are pointed to the references mentioned for complete details.

The observable in the global 21~cm experiments is the cosmic mean differential brightness temperature, $\delta T_b(\nu)$, which is measured relative to the cosmic microwave background (CMB) temperature and can be expressed as \citep{furlanetto2006c} 
\begin{equation}
  \delta T_b (\nu) = \frac{T_S(z)-T_\gamma(z)}{1+z}\left(1-e^{-\tau} \right),
 \end{equation} 
where $\tau$ is the 21~cm optical depth of the diffuse IGM and  $T_{S}$ and $T_{\gamma}$ are the neutral hydrogen (\HI) spin temperature and the background radiation temperature, respectively.
%
%
%
%
The optical depth is usually much less than unity. In this case, for the cosmological parameters used in the paper, the above expression simplifies to 
\begin{equation}
    \delta T_b(\nu) \approx 10.1~\mbox{mK}~\chi_{\rm HI}(z) \left(1 - \frac{T_{\gamma}(z)}{T_S(z)}\right)~(1+z)^{1/2},
\label{eq:delta_Tb}
\end{equation}
where $\chi_{\mathrm{HI}}$ is the neutral hydrogen fraction.

In absence of any other radiation sources at radio frequencies $\sim 1420$~MHz, the radiation temperature $T_{\gamma}(z)$ is given by the CMB temperature $T_{\rm CMB}(z)$. The spin temperature $T_S$ can be written as a weighted sum of various temperatures \citep{field1958} such that
\begin{equation}
    T_S^{-1}=\frac{T_{\gamma}^{-1}+x_cT_{k}^{-1}+x_{\alpha}T_\alpha^{-1}}{1+x_c+x_{\alpha}},
\end{equation}
where $T_k$ is the gas kinetic temperature and $T_{\alpha}$ is the color temperature of the Ly$\alpha$ radiation field. Further, $x_c$ and $x_{\alpha}$ are the coupling coefficients corresponding to the collisional excitations and spin-flip due to the Ly$\alpha$ radiation field, respectively. For all practical purposes one can assume $T_{\alpha}=T_k$. This is justified by the fact that the optical depth for Ly$\alpha$ photons is so high that they undergo a large number of scatterings -- these are sufficient to bring the Ly$\alpha$ radiation field and the gas into local equilibrium near the central frequency of Ly$\alpha$ radiation \citep{pritchard2012}. 

We now discuss how the coupling coefficients and the gas kinetic temperature are calculated.
\subsection{The coupling co-efficients}
The coupling coefficients used in this work are calculated as follows: 

{\it (i)} The collisional coupling coefficient, $x_c$, is determined by three different channels, namely, hydrogen-hydrogen (H-H), hydrogen-electron (H-e) and hydrogen-proton (H-p) collisions. The results in a total coupling coefficient of
\begin{equation}
x_c = \frac{T_*}{A_{10}T_{\gamma}} \sum_{i = {\rm H, e, p}} \kappa_{10}^{{\rm H}-i} T_{k} n_i,
\end{equation}
where $T_* = 68.5~\mbox{mK}$ is the temperature corresponding to the 21~cm transition, $A_{10}$ is the corresponding Einstein-coefficient for spontaneous emission and $\kappa_{10}^{{\rm H}-i}$ is the specific rate coefficient for spin de-excitation by collisions with particles of species $i$ with number density $n_i$. The collisional coefficients $\kappa_{10}^{{\rm H}-i}$ do not play any important role at epochs relevant to this work, however, for completeness, we include them in our numerical code using the fitting functions given in \citet{kuhlen2006} and \citet{liszt2001}.

{\it (ii)} The  Ly$\alpha$ coupling coefficient, $x_{\alpha}$, also known as the Wouthuysen-Field coupling coefficient, is essentially determined by the background Ly$\alpha$ flux through the following relation \citep{furlanetto2006b}
\begin{equation}
    x_{\alpha}=1.81 \times 10^{11} (1+z)^{-1} S_{\alpha} \frac{J_{\alpha}} {\mbox{cm}^{-2} \mbox{s}^{-1} \mbox{Hz}^{-1} \mbox{sr}^{-1}},
\end{equation}
where $S_{\alpha}$ is a factor of order unity that accounts for the detailed atomic physics involved in the scattering process \citep{furlanetto2006b, hirata2006}. Further, $J_{\alpha}$, the background Ly$\alpha$ photon flux, is calculated using the {\it Delphi} model as \citep{2003ApJ...596....1C}
\begin{equation}
J_{\alpha}(z)=\frac{c(1+z)^{3}}{4\pi}\int_z^{\infty} \dot{n}_{\nu'}(z') \left|\frac{\der t'}{\der z'}\right|  \der z',
\end{equation}
where $\nu' = \nu_{\alpha} (1 + z') / (1 + z)$ with $\nu_{\alpha}$ being the Ly$\alpha$ frequency, $\dot{n}_{\nu'}(z')$ is the production rate of photons per unit frequency per unit comoving volume at redshift $z'$, $c$ is the speed of light and $t'$ is the cosmic time corresponding to the redshift $z'$.  In this paper, we assume the spectrum of photons around the Ly$\alpha$ frequencies to be constant \citep{furlanetto2006b}. 

\subsection{The gas kinetic temperature}
The next quantity of interest is the gas kinetic temperature which evolves as \citep{furlanetto2006b}
\begin{equation}
    \frac{\der T_k}{\der z} = \frac{2 T_k}{1+z}-\frac{2}{3H(z)(1+z)}{\sum}_{i}\frac{\epsilon_i}{k_B n}.
\end{equation}
The first term on the right hand side corresponds to the cooling due to the adiabatic expansion of the Universe while the second term accounts for the heating and cooling processes summarised below. The quantity $\epsilon_i$ is the energy injected into (or taken out from) the gas per second per unit physical volume through process $i$, $n$ is the total number of gas particles and $k_B$ is Boltzmann's constant. The key heating and cooling processes relevant at high-$z$ include:

{\it (i) Compton heating/cooling}: The rate of heating/cooling due to Compton scattering of residual electrons with background photons is given by
\begin{equation}
    \frac{2\epsilon_{comp}}{3nk_B}=\frac{\chi_e}{1+\chi_e+f_{He}}\frac{8\sigma_T u_{\gamma}}{3m_e c}\left( T_{\gamma}-T_{k}\right)
\end{equation}
where $\chi_e=(1-\chi_{HI})$ is the ionization fraction \citep[$\chi_e\simeq 10^{-4}$;][]{bharadwaj2004}, $u_{\gamma}$ is the energy density of background photons, $\sigma_T$ is the Thomson cross-section, $f_{He}=0.08$ is the helium fraction by number and $m_e$ is the electron mass.

{\it (ii) X-ray heating/cooling}: The X-ray photons produced by early sources (e.g., accreting black holes, miniquasars, supernova shocks or X-ray binaries) can heat up gas. The amount of heating depends both on the number of photons produced as well as the spectral shape. Given these quantities are highly uncertain at high-$z$, they are almost impossible to model in a self-consistent manner. In this work, taking a somewhat conservative and simple approach, we assume that the correlation between $\dot{M}_*$ and the X-ray luminosity, $L_X$, of galaxies observed in the local Universe \citep{grimm2003} also holds at high-$z$. We parametrize this correlation as \citep{furlanetto2006b}
\begin{equation}
    L_X=3.4 \times 10^{33} f_X \left(\frac{\dot{M}_*}{\text{M}_{\odot}~\text{yr}^{-1}} \right) \mbox{J~s}^{-1},
\end{equation}
where $f_X$ is an unknown normalization factor allowing one to account for differences between local and high-$z$ observations. Note that the more recent data indicate a different value for the normalization of the $\dot{M}_* - L_X$ relation \citep{2012MNRAS.419.2095M}, however, any difference in the normalization is completely absorbed in the free parameter $f_X$. We hence use the value used by \citet{furlanetto2006b} so as to allow easy comparison with their results. Given the $\dot{M}_*$ values yielded by the {\it Delphi} model, we calculate the star formation rate density (SFRD; $\dot{\rho}_*$). The globally averaged energy injection rate per unit volume can then be expressed as
\begin{equation}
  \epsilon_X =3.4 \times 10^{33} f_h~f_X~\frac{\dot{\rho}_*}{\text{M}_{\odot}~\text{yr}^{-1}~\text{Mpc}^{-3}}{\rm J ~s^{-1} Mpc^{-3}},
\end{equation}
where $f_h$ is the fraction of the X-rays that contribute to heating (the other part goes into ionization). We combine $f_X$ and $f_h$ into one free parameter $f_{X,h} = f_X \times f_h$. 

Finally, we ignore Ly$\alpha$ heating in our work as it is believed to be less important compared to X-ray heating \citep{chen2004}. We also ignore the heating and cooling processes during the reionization epoch ($z \lesssim 15$) since, by that time, the IGM is heated up to temperatures well above the CMB, and hence the 21~cm signal becomes insensitive to the exact value of the spin temperature.


\begin{figure*}
	\centering
	\begin{subfigure}{0.4\textwidth} 
		\includegraphics[width=1.1\textwidth]{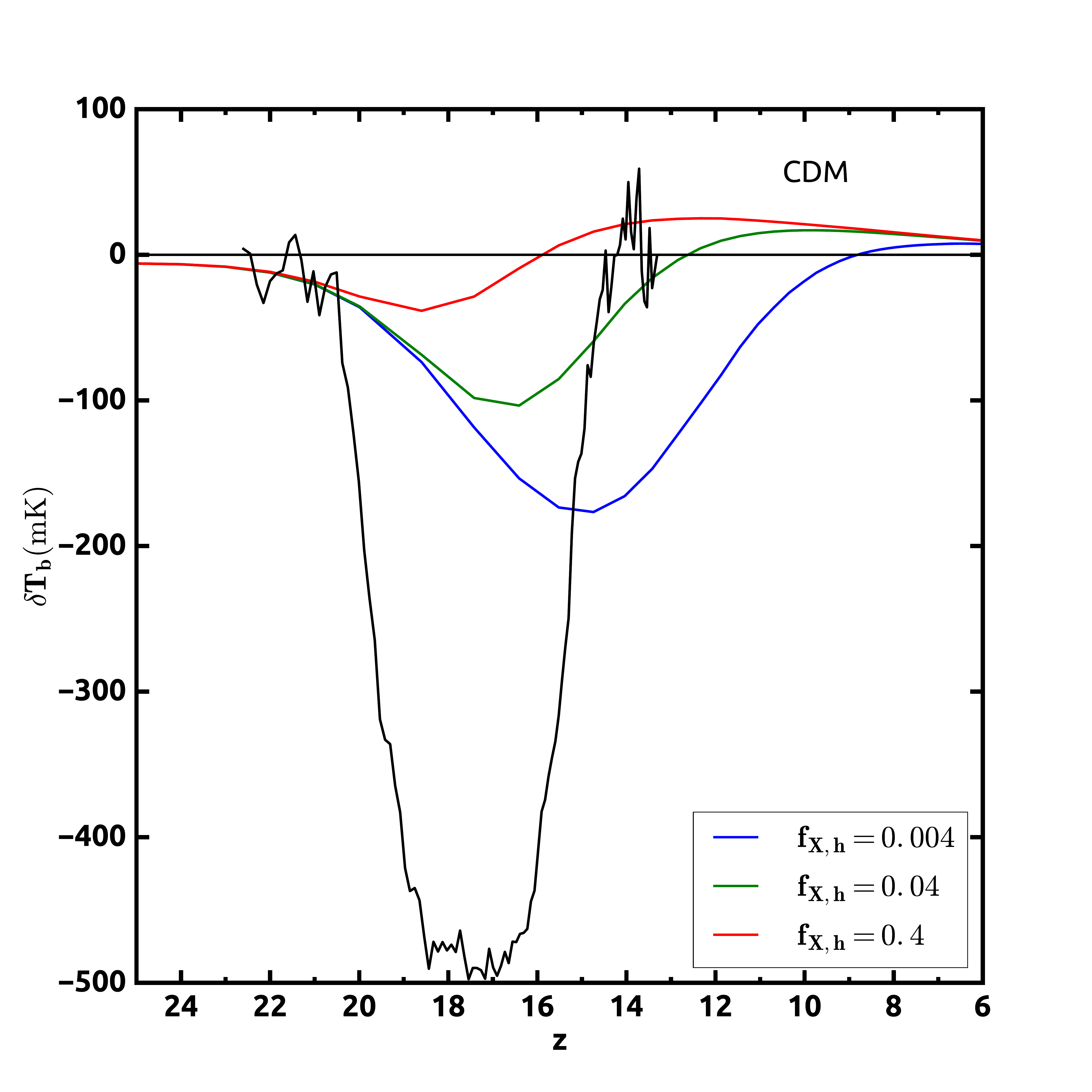}
	\end{subfigure}
	\begin{subfigure}{0.4\textwidth} 
		\includegraphics[width=1.1\textwidth]{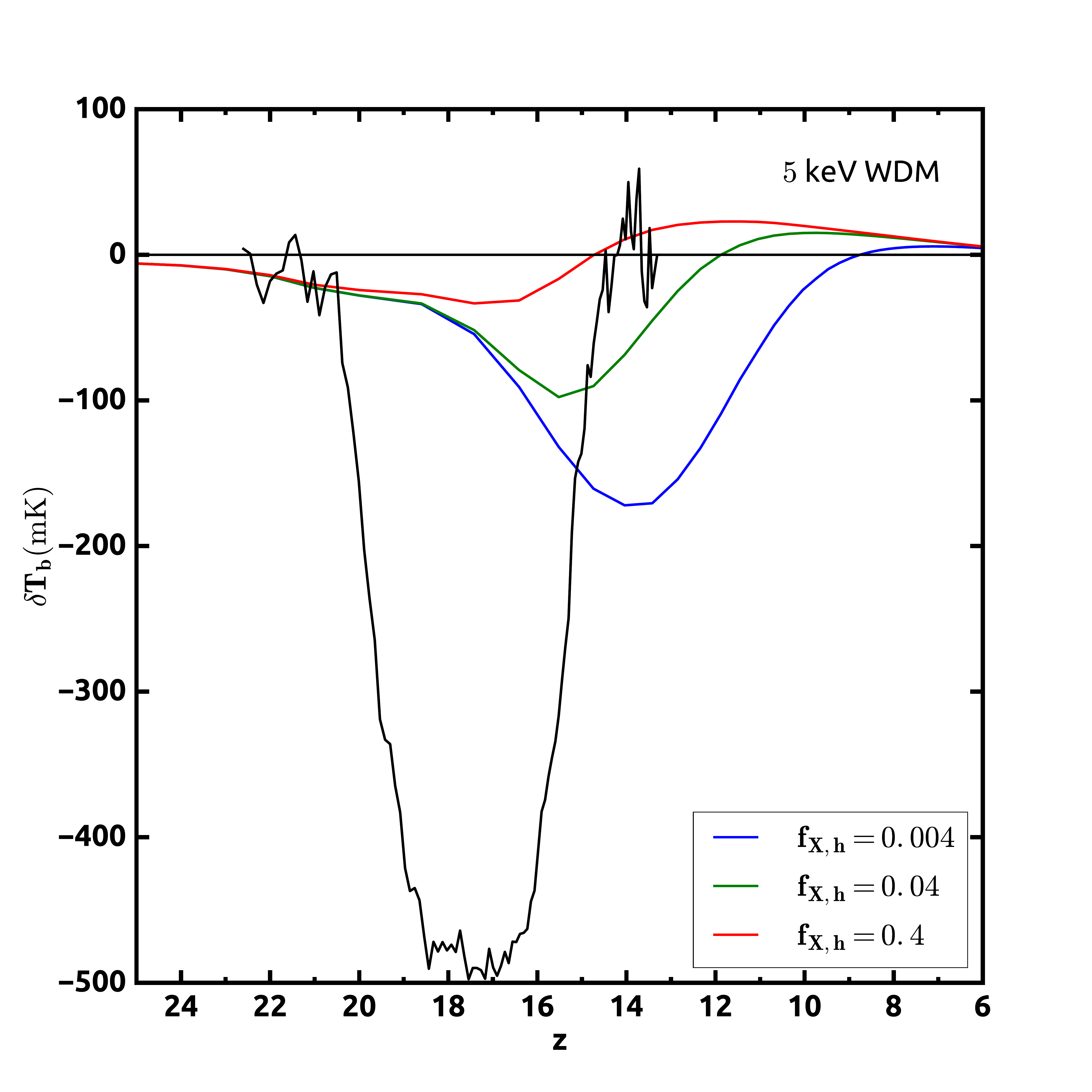}
	\end{subfigure}
	\begin{subfigure}{0.4\textwidth} 
		\includegraphics[width=1.1\textwidth]{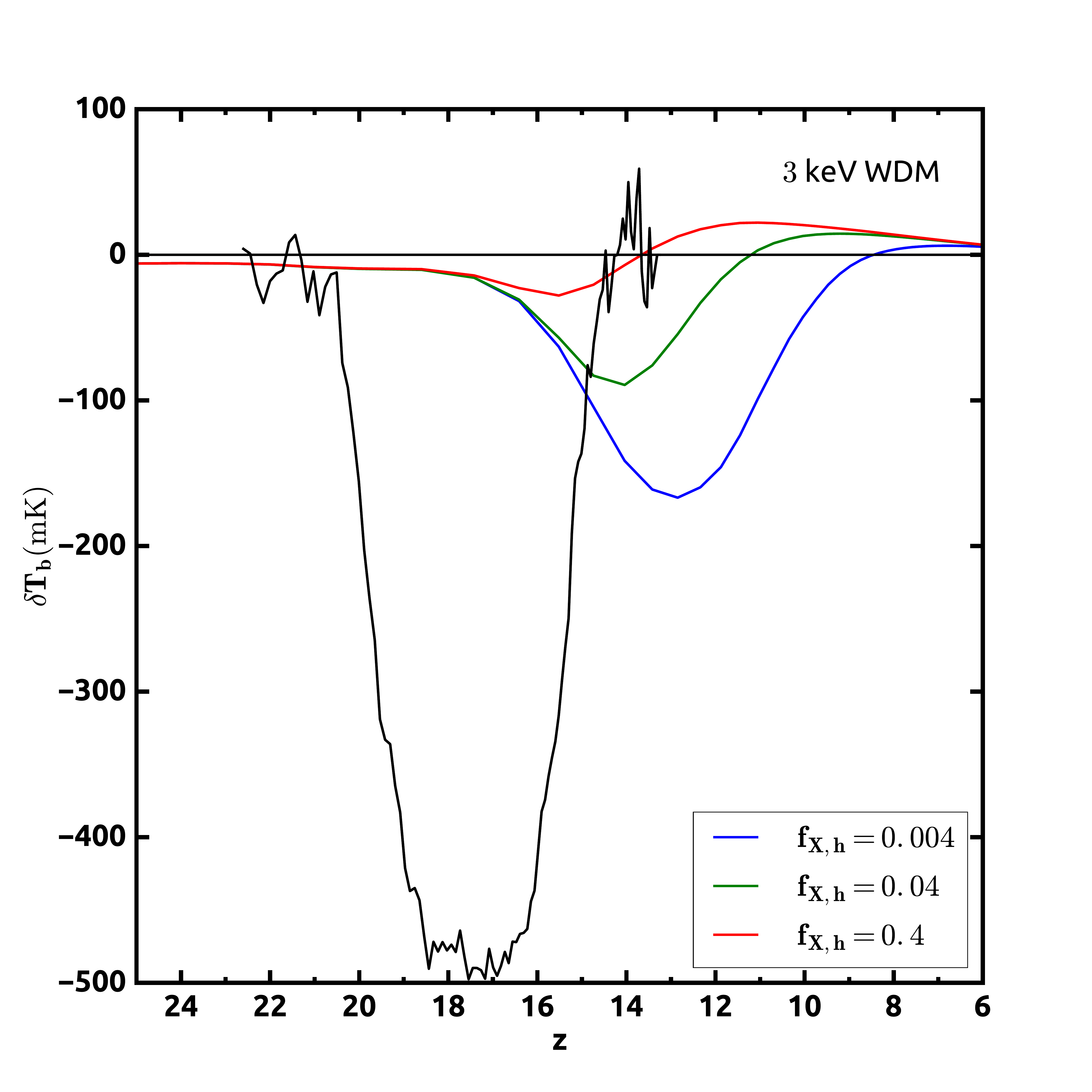}
	\end{subfigure}
	\begin{subfigure}{0.4\textwidth} 
		\includegraphics[width=1.1\textwidth]{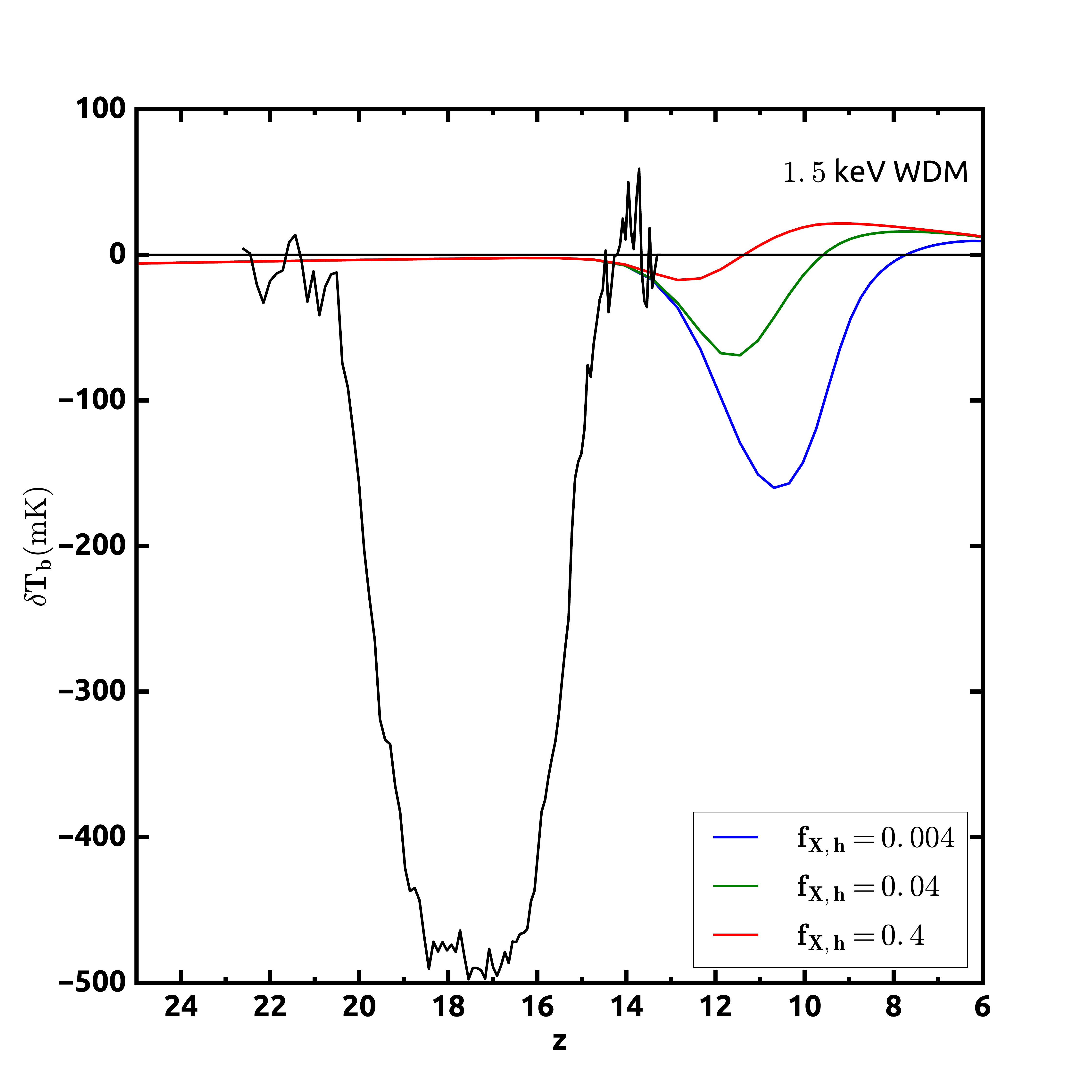}
	\end{subfigure}
	\hspace{1em} 
	\caption{The global 21~cm differential brightness temperature for the CDM, 5 keV, 3 keV and 1.5 keV WDM models, as marked, including X-ray heating only (see Sec. \ref{xray} for details). In each panel the black curve shows the EDGES result; as marked, the other coloured lines show our model predictions for the different values of $f_{X,h}$ related to X-ray heating at these high-$z$. As shown, while some models lie within the redshift range of the EDGES signal, none match the amplitude of the observed signal.} 
\label{fig:default_4_panels} 
\end{figure*}

Although the 21~cm signal relevant for comparing with the EDGES data is relatively insensitive to the reionization history, we still compute it for completeness. We use the production rate of ionizing photons produced per unit time per unit comoving volume, $\dot n_{ion}$, obtained from {\it Delphi} for each galaxy, to compute the evolution of the global \HI\, fraction as
\begin{equation}
\frac{\der \chi_{\rm HI}}{\der t} = -f_{\rm esc}\frac{\dot{n}_{\rm ion}}{n_{H, {\rm com}}} + (1 - \chi_{\rm HI})~ \alpha_{B}~{\cal C}~n_{H,{\rm com}}~(1+z)^{3},
\end{equation}
where $n_{H, {\rm com}}$ is the hydrogen comoving number density, $f_{\rm esc}$ is the escape fraction of ionizing photons, $\alpha_B$ is the (case B) recombination rate coefficient and ${\cal C}$ is the clumping factor of the IGM.

At this point, our model has three free parameters: $f_{X,h}$, $f_{\rm esc}$ and ${\cal C}$. While the first is relevant for the cosmic dawn, the latter two are crucial for the reionization history. As shown in our previous works \citep{dayal2017a}, both the CMB optical depth $\tau_{esc}=0.055\pm 0.009$ \citep{planck2016} and ionizing emissivity constraints at $z \gsim 6$\footnote{The emissivity is calculated using the approach outlined in \citet{2012MNRAS.423..862K}, i.e., by combining the observational constraints on the hydrogen photoionization rate from \citet{2011MNRAS.412.1926W} and the mean free path of ionizing photons from \citet{2010ApJ...721.1448S}.} can be simultaneously fit, for all four dark matter models, using ${\cal C} =1+43z^{-1.71}$ \citep{pawlik2009, haardt2012} and $f_{\rm esc}$ that evolves as
\begin{equation}
f_{\mathrm{esc}}(z)=\mathrm{min} \left[1,f_{0}\left(\frac{1+z}{7} \right)^{\alpha} \right] \, \mathrm{for}\, z\ge5.
\end{equation}
Here, $f_0\times 100 = 4.5\, (4.1, 3.8, 4.8)$ and $\alpha = 2.9\, (3.7, 4.3, 6.2)$ for the CDM (5 keV, 3 KeV and 1.5 keV WDM) model.

\section{EDGES constraints on the WDM mass}

We now present our results for the global 21~cm signal during cosmic dawn for the different dark matter models considered for two scenarios: the first where only X-ray heating is accounted for and the second where we include an excess radio background along with the X-ray heating. 

\subsection{Models with X-ray heating}
\label{xray}
In the first case where only X-ray heating is considered, the 21~cm differential brightness temperatures for the four dark matter models considered, along with the EDGES observations, are shown in \fig{fig:default_4_panels}. For each dark matter model, we start by showing results for typical values of $f_{X}=0.2$ \citep{glover2003} and $f_{h}=0.2$ \citep{furlanetto2006b}, yielding $f_{X,h} = 0.04$. We then carry out calculations varying $f_{X,h}$ by an order-of-magnitude on either side to see how this impacts our results. 

We find that, independent of the dark matter model used, as $f_{X,h}$ increases, the location of the maximum absorption shifts to higher redshifts and has a smaller amplitude. This is because the higher the value of $f_{X,h}$, the earlier the IGM begins heating up, shifting the absorption signal to higher-$z$ as well as leading to a decrease in its amplitude. However, we note that the redshift at which the absorption signature begins to show up corresponds to the formation of the first galaxies in our models and is thus independent of the X-ray heating efficiency. Given the progressive delay in structure formation going from cold to warm dark matter models, this results in the absorption signal appearing at progressively later redshifts from CDM to WDM with decreasing $m_x$ values. Further, the faster build-up of stellar mass (and hence the SFR) with decreasing $m_x$ \citep[see e.g.][]{dayal2015} results in a shorter time interval between the onset of star formation and efficient X-ray heating - this naturally decreases the amplitude of the absorption profile for decreasing $m_x$ values as compared to CDM. 

Finally, it is clear that, irrespective of the underlying cosmology, none of the models discussed above produce an absorption signal with an amplitude larger than about $-180$ mK which is much smaller than the amplitude of $-500\pm 200$mK measured by the EDGES observations. As noted above, this result is consistent with expectations \citep[e.g.][]{barkana2018} and requires additional physics to be incorporated into models \citep{barkana2018, fraser2018, pospelov2018, slatyer2018}. We can also take a step back and relax the constraint on the amplitude, only demanding that the absorption signal is fully contained between $13 \lsim z \lsim 21$ so as to be compatible with the EDGES observations \citep[e.g.][]{2018PhRvD..98f3021S}. As shown in (panels a and b of) \fig{fig:default_4_panels}, we find that the CDM  and 5keV WDM models produce such a signal for $f_{X,h} \gsim 0.04$ and $f_{X,h} \gsim 0.4$, respectively. However, given the delay in structure formation, $m_x \lsim 3$ keV WDM models are unable to produce a signal in the observed redshift range, irrespective of the $f_{X,h}$ values used. Thus, even with this first estimate, the EDGES signal can be used to {\it rule out $m_x\lsim 3$ keV WDM.}

\subsection{Models with excess radio background}
\label{excess}
As seen from the above section, matching the amplitude of the EDGES signal would either require making the gas colder, using exotic physics such as new dark matter interactions \citep{barkana2018, fraser2018, pospelov2018, slatyer2018} or invoking a radiation temperature larger than the CMB temperature \citep{fraser2018,pospelov2018,ewall_wice2018,feng2018}. In this work, we avoid incorporating any non-standard physics and focus on the latter scenario. Such an approach is lent some support by observations of an excess radio background in the local Universe as reported by the ARCADE-2 experiment \citep{fixsen2011}. 

We model the excess radio background by assuming that early galaxies produce radio frequency radiation whose strength is proportional to the SFR. The local radio-SFR ($L_{R}-\dot{M}_*$) relation at $150$ MHz is given by \citet{gurkan2018}
\begin{equation}
L_R=f_{R} \times 10^{22} \left(\frac{\dot{M}_*}{1\text{M}_{\odot}\text{yr}^{-1}}\right)\mbox{J~s}^{-1}\mbox{Hz}^{-1}
\end{equation}
where $f_R$ is a free parameter. We extrapolate this relation to higher frequencies by assuming a spectral index of $-0.7$ \citep{gurkan2018}. The globally averaged radio luminosity per unit comoving volume (i.e., the radio emissivity) at redshift $z$ is then given by
\begin{equation}
  \epsilon_R(z) = f_R \times 10^{22}\times \frac{\dot{\rho}_*(z)}{\text{M}_{\odot}~\text{yr}^{-1}~\text{Mpc}^{-3}}{\rm J s^{-1} Hz^{-1} Mpc^{-3}}.
  \end{equation} 
The corresponding 21~cm radiation flux at a redshift $z$ from such high-$z$ sources can then be written as \citep{2003ApJ...596....1C}
\begin{equation}
\begin{aligned}
F_{R}(z)= &  \left(\frac{1420}{150}\right)^{-0.7} \frac{c(1+z)^{3}}{4\pi} \int_z^{\infty} \epsilon_{R, \nu'}(z') \left|\frac{\der t'}{\der z'} \right| \der z',
\end{aligned}
\end{equation}
where $\epsilon_{R, \nu'}(z')$ is the comoving radio emissivity at $\nu' = 150~{\rm MHz} (1 + z') / (1 + z)$. We convert this flux into a radio brightness temperature $T_{R}$. This results in a total background temperature given by $T_{\gamma}(z)=T_{R}(z)+T_{\rm CMB}(z)$. 

\begin{figure*}
	\centering
	\begin{subfigure}{0.4\textwidth} 
		\includegraphics[width=1.1\textwidth]{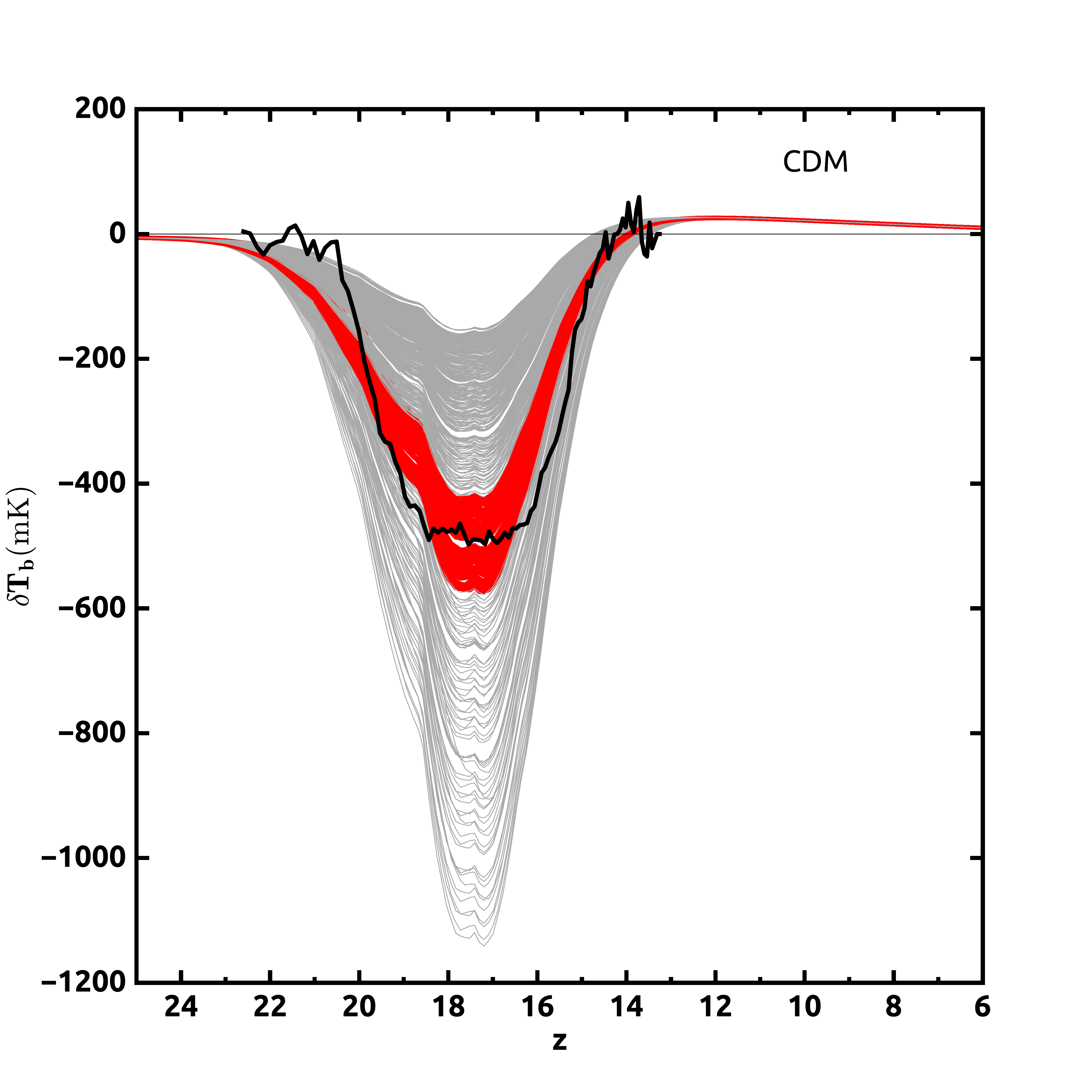}
	\end{subfigure}
	\begin{subfigure}{0.4\textwidth} 
		\includegraphics[width=1.1\textwidth]{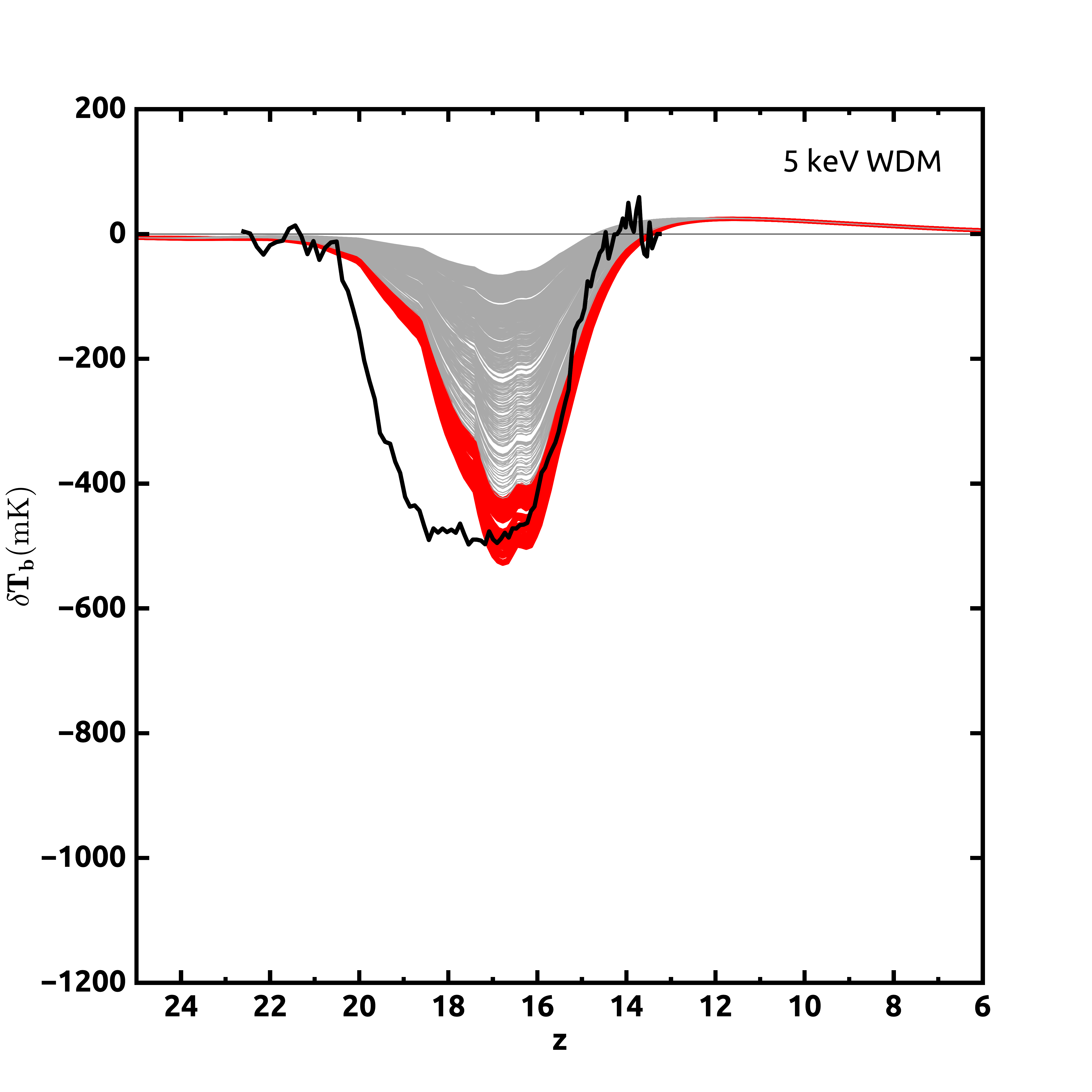}
	\end{subfigure}
	\begin{subfigure}{0.4\textwidth} 
		\includegraphics[width=1.1\textwidth]{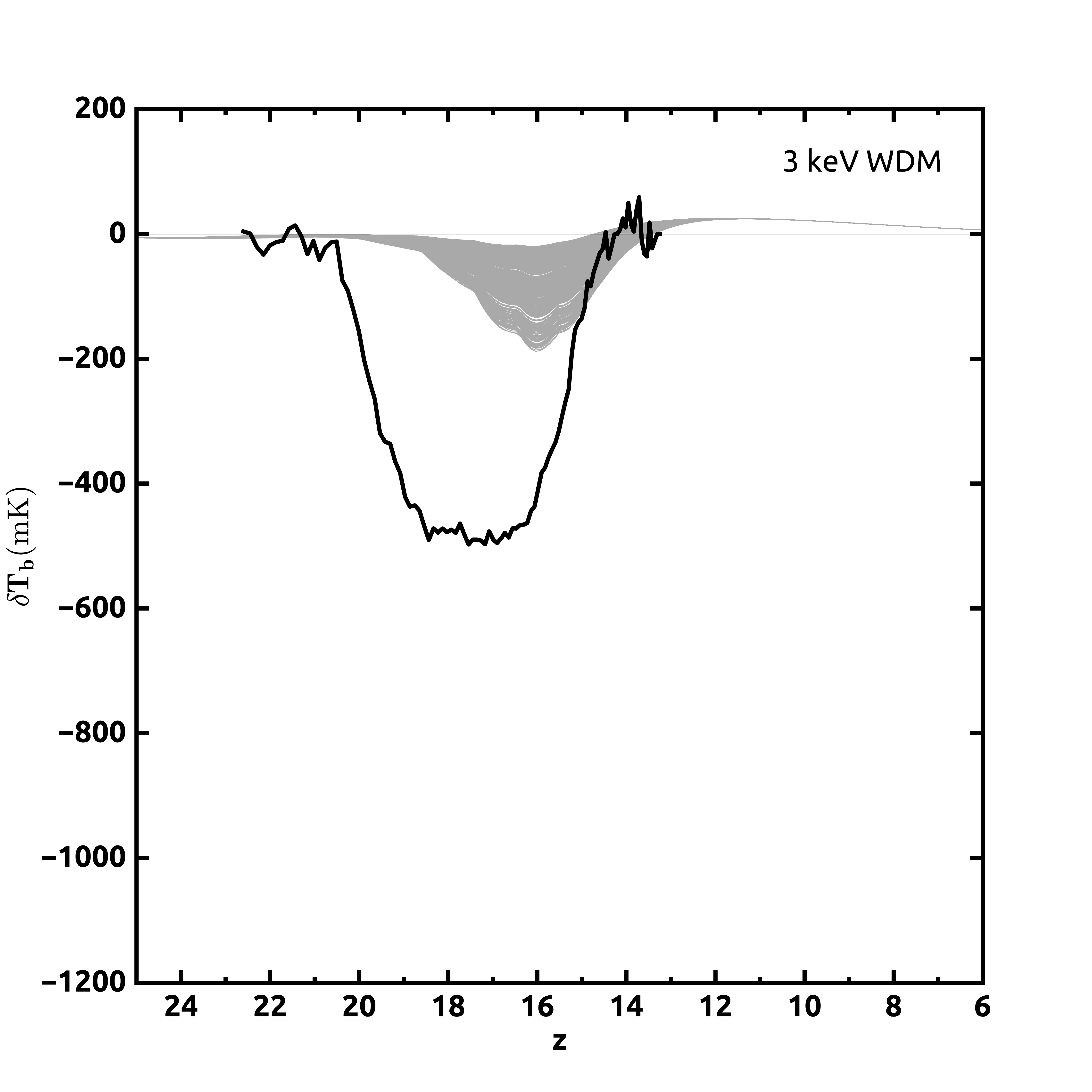}
	\end{subfigure}
	\begin{subfigure}{0.4\textwidth} 
		\includegraphics[width=1.1\textwidth]{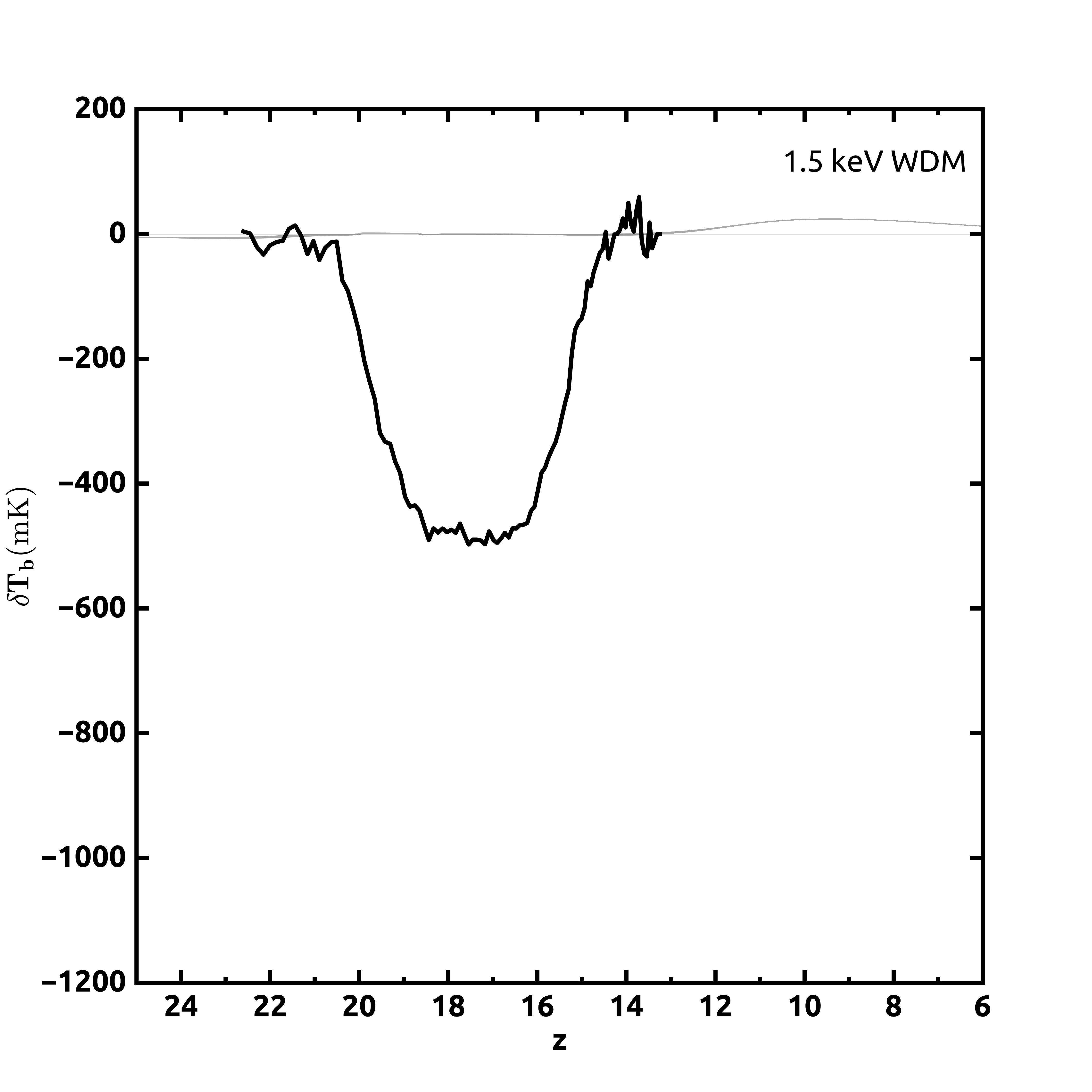}
	\end{subfigure}
	\caption{The global 21~cm differential brightness temperature for the CDM, 5 keV, 3 keV and 1.5 keV WDM models, as marked, including both X-ray heating and an excess radio background (see Sec. \ref{excess}). In each panel the black curve shows the EDGES result. The grey lines in each panel show models that satisfy the ARCADE-2 limits and where the signal is limited to $z \gsim 14$. The red lines show models consistent with the EDGES result, both in terms of the redshift range of the signal as well as its amplitude ($\delta T_b = -500\pm 75$mK). As shown, the inclusion of an excess radio background results in free parameter combinations ($f_R$ and $f_{X,h}$) yielding results in agreement with the EDGES data for the CDM and 5 keV WDM models. However, $m_x \lsim 3$ keV models can effectively be ruled out since they are unable to reproduce either the redshift range or the  amplitude of the observed signal.}
\label{fig:excess_radio} 
\end{figure*}

We then calculate the 21~cm differential brightness temperatures over a two-dimensional grid in $f_R = 10^{3-11}$ and $f_{X, h} = 10^{0-7}$ for CDM; increasingly light WDM models require increasing values of both these parameters for which the final fine-grid values explored are cosmology-dependent. While using $f_R$ can indeed enhance the amplitude of the absorption signal, a redshift independent value leads to an enhancement that is more extended in redshift-space than the EDGES signal. Inspired by the results of \citet{mirocha2019}, we turn off this excess radio background at $z = 16$. In order to ensure numerical stability of the code, we model the radio background as a tanh function having a width $\Delta z \approx 1.5$; our results are insensitive to the precise choice of the width. Further, we reject all combinations which produce a radio background higher than that observed by ARCADE-2 at $z=0$. Note that this is a conservative choice as low-$z$ galaxies are expected to produce an additional radio signal not accounted for in this work. We find that in such a formalism a variety of combinations of $f_{X,h}$ and $f_R$ can match the EDGES signal in redshift and amplitude as shown in \fig{fig:excess_radio}. For the CDM and 5 keV models, we show only those models that simultaneously satisfy the ARCADE-2 upper limits and where the absorption signal is strictly limited to $z \gsim 14$. However, as shown in the same figure, {\it there are no combinations of free parameters that can reproduce the amplitude ($-500\pm75$mK), or even a signal in the required redshift range, for 3 keV and 1.5 keV WDM models given the delay in galaxy formation in these models. Based on the EDGES and ARCADE-2 observations, we can therefore rule out WDM model with $m_{X}\leq 3$ keV.}

 While the presence of the additional radio background during cosmic dawn might be expected because of the formation of the first stars\footnote{However, there are concerns in sustaining the radio background produced by accelerating relativistic electrons because their cooling time-scale is much shorter than the Hubble time \citep[see, e.g.,][]{2018MNRAS.481L...6S}.}, the physical reason for this background being suppressed at $z \lsim 16$ is much more difficult to explain. One possible reason could be the radio background being mostly powered by PopIII stars which tend to die off as the gas gets metal-enriched. The key issues with such an explanation, however, are the quick transition from metal-free to metal-enriched star formation required as well as the fact that most models of  PopIII star formation predict such a transition at much lower redshifts. Another possibility is that the production of a radio signal from the first stars could immediately be followed by heating of gas by cosmic rays \citep{jana2019}. Since the signal depends only on the ratio $T_{\gamma} / T_S$, this heating will have the same effect as decreasing the radiation temperature. While we cannot provide a fully self-consistent model for the radio background at the moment, it is clear that explaining the EDGES data (without incorporating any exotic physics) requires an excess radio background during cosmic dawn that switches off at $z \sim 16$.

\begin{figure*}
\center{\includegraphics[scale=0.75]{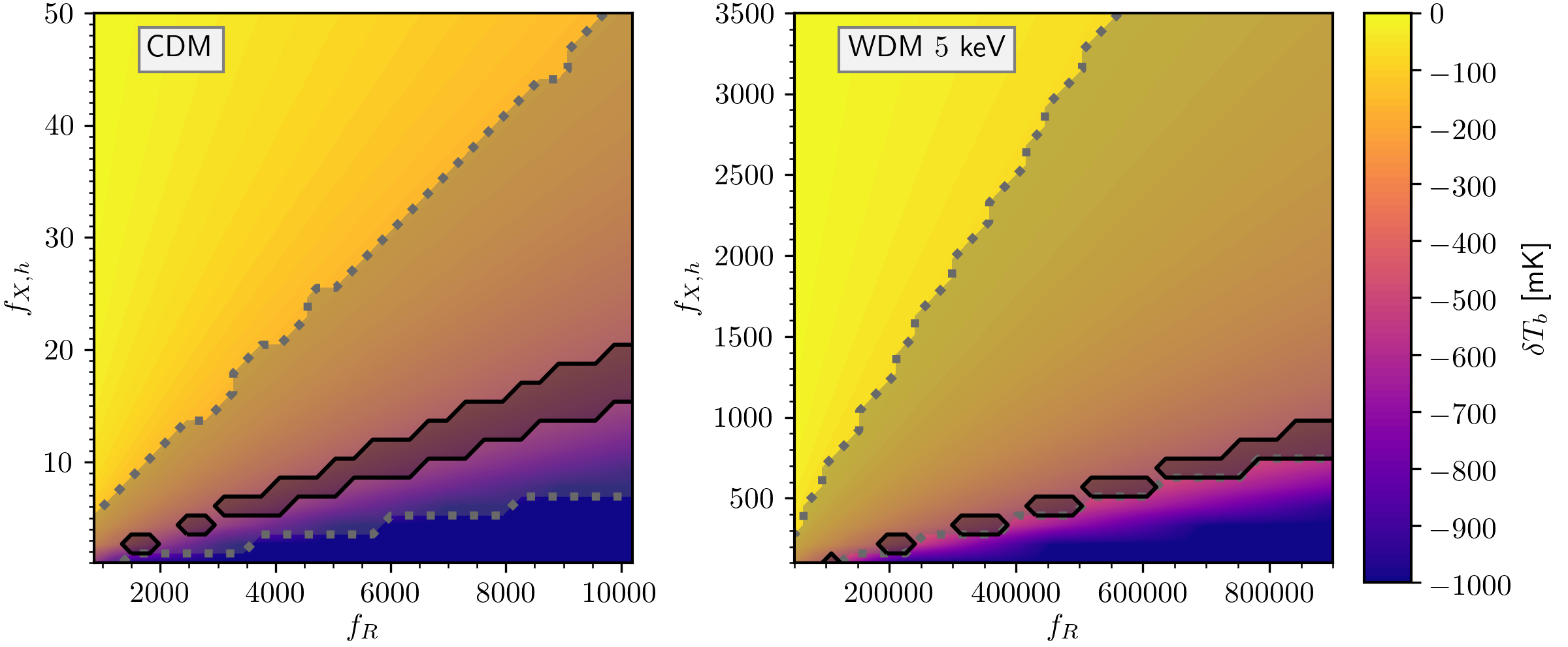}}
\caption{The differential brightness temperature, color-coded as per the color bar, for different combinations of the two model free parameters: $f_{X,h}$ and $f_R$ that account for X-ray heating and an excess radio background respectively for CDM ({\it left panel}) and 5 keV WDM ({\it right panel}). The shaded area demarcated by the dot-dashed lines shows parameter combinations in accord with ARCADE-2 results and the EDGES redshift range. The dark-shaded areas show parameter combinations that, additionally, match the brightness temperature measured by EDGES ($-500\pm 75$mK).  }
\label{fig3} 
\end{figure*}

Finally, we show the differential brightness temperature over a two-dimensional grid composed of the two model free parameters in Fig. \ref{fig3}. We demarcate two regions: the first (light-shaded; bounded by dot-dashed lines) that reproduces ARCADE-2 results as well as a 21~cm signal in the redshift range measured by EDGES and the second (dark shaded; bounded by solid lines) that, additionally, matches the $\delta T_b = -500 \pm 75$ mK signal measured by EDGES. As shown, an increase in $f_{X,h}$ that increases X-ray heating of gas leading to a shallower dip, must be compensated by an increasing $f_R$ value that increases the radio background, increasing the depth of the 21~cm signal. As expected, the later emergence of structure in the 5 keV model, as compared to CDM,  requires larger values of both these parameters to yield a similar 21~cm signal.  

\section{Conclusions and discussion}
This work has focused on constraining the warm dark matter particle mass by comparing results of our galaxy formation model, {\it Delphi} \citep{dayal2014, dayal2015}, with the global 21~cm signal recently detected by the EDGES collaboration \citep{bowman2018}. Our work is based on the fact that the frequency of the EDGES signal implies the first stars to have formed as early as $z \sim 18$. The increasing delay in galaxy formation in progressively light WDM models could therefore be used to constrain the WDM mass at these early epochs, inaccessible by any other means.

Starting from a scenario wherein the radiation temperature is solely provided by the CMB and in absence of any non-standard cooling of the gas, none of our models can match the strength of the absorption signal measured by EDGES. Working with less stringent criteria and only demanding that the models predict the absorption signal at the redshift location demarcated by EDGES, we can essentially rule out $m_x \lsim 3$ keV WDM given the delay in galaxy formation, and hence the build-up of the Ly$\alpha$ background, in such models.

We also propose a way to reproduce both the redshift and amplitude of the EDGES signal by introducing an additional source of radio background radiation which scales with the SFRD. Some indication of such an excess exists from local observations with ARCADE-2 \citep{fixsen2011}. However, we find that a redshift-independent scaling between the radio background and the SFRD results in an absorption that extends much later than conventional models as well as the EDGES signal. Reproducing the EDGES data requires such a background be turned off (or equivalently, the gas be heated up beyond what is predicted by X-ray heating) at $z \sim 16$. As of now, both the sources (metal free stars or cosmic rays associated with the first stars and/or black holes) and reasons for the decay of such a background at $z <16$ remain open questions. Despite this and irrespective of the values of the free parameters used, in this case too, structure formation is delayed long enough in $m_x \lsim 3$ keV WDM models so that they can be ruled out. 

It is worth pointing out here that our galaxy formation model is calibrated to data at relatively lower redshifts $z \lesssim 10$. It is possible that the star formation during the cosmic dawn is dominated by processes different from what have been included here. However, it would still be difficult to produce an absorption signal consistent with EDGES for WDM models with $m_x\lsim 3$ keV as there are almost no dark matter haloes at these high redshifts. In addition, our calculations of the 21~cm signal contains various simplifying assumptions at different stages (e.g., the various efficiency parameters related to X-ray heating are taken to be independent of $z$). This too will not affect our conclusions which we have checked by varying the free parameters to their extreme limits. It is clear that {\it WDM models with $m_x\lsim 3$ keV simply cannot form stars early enough to satisfy the EDGES constraint}.
Our constraints are thus comparable to constraints obtained from the Ly$\alpha$ forest data \citep{viel2005,baur2016,irsic2017} and are crucial in that they extend this constraint on the WDM particle mass to the first 200 million years of the Universe.

Finally, another issue, related to the observations, is that there might be uncertainties regarding the fitting of the foregrounds in the EDGES experiment \citep{2018Natur.564E..32H}. In that case, the signal may not be what is reported by \citet{bowman2018}. While exploring this is beyond the scope of our work, results from other ongoing experiments that aim at detecting the global 21~cm signal will provide a crucial avenue to using astrophysical observations to shed light on the nature of dark matter.

\section{Acknowledgments} 
TRC acknowledges support from the Associateship Scheme of ICTP, Trieste. PD and AH acknowledge support from the European Research Council's starting grant DELPHI (717001). PD also acknowledges support from the European Commission's and University of Groningen's CO-FUND Rosalind Franklin program. 

\bibliographystyle{mn2e}
\bibliography{edges_rv}

\end{document}